\documentclass[preprint,5p,times,twocolumn]{elsarticle}

\usepackage{amssymb}
\usepackage{lipsum}

\usepackage{url} 
\usepackage{lineno}
\usepackage{soul}

\usepackage{graphicx}
\usepackage{bm}
\usepackage{booktabs}
\usepackage{array}
\usepackage{cleveref}
\usepackage{tikz}

\usepackage{xcolor, soul}
\sethlcolor{black}

\usepackage{float}

\usepackage{fancyhdr}

\usepackage{verbatim}

\UseRawInputEncoding
\usepackage[pscoord]{eso-pic}

\newcommand{\placetextbox}[3]{
  \setbox0=\hbox{#3}
  \AddToShipoutPictureFG*{
    \put(\LenToUnit{#1\paperwidth},\LenToUnit{#2\paperheight}){\vtop{{\null}\makebox[0pt][c]{#3}}}%
  }%
}%

\usepackage{lineno}

\journal{Nuclear Physics A}

\begin{document}

\begin{frontmatter}



\title{The iSTORM Instrument for Airborne Measurements of Gamma-Ray Emissions from Thunderstorms}


\author[1]{Daniel Shy}
\author[1]{J. Eric Grove}
\author[1]{Bernard Phlips}
\author[1]{Alena Schell}
\author[1]{Mary Johnson-Rambert}

\author[2]{Mason Quick}
\author[3]{David Corredor}
\author[3]{Scott Podgorny}

\author[4]{Roy Salinas}
\author[1]{Mitch Davis}

\address[1]{U.S. Naval Research Laboratory, 4555 Overlook Ave. SW, Washington, DC, USA}

\address[2]{NASA Marshall Space Flight Center, Huntsville, AL, USA}
	      
\address[3]{Earth Systems Science Center, University of Alabama in Huntsville, Huntsville, AL, USA}

\address[4]{National Research Council Research Associate resident at the U.S. Naval Research Laboratory, 4555 Overlook Ave., SW, Washington, DC, USA}

\begin{abstract}

	The in-Situ Thunderstorm Observer for Radiation Mechanisms (iSTORM) is a gamma-ray spectrometer designed to study gamma-ray transients originating from thunderstorms, such as glows and terrestrial gamma-ray flashes (TGFs). It is designed and built by the U.S. Naval Research Laboratory for deployment aboard a NASA ER-2 aircraft. Using an array of 32 one-inch-diameter $\mathrm{CeBr}_3$ scintillators read out with silicon photomultipliers (SiPMs), the instrument achieves an energy range of $\sim 250 \ \mathrm{keV}$ to $5 \ \mathrm{MeV}$ under flight conditions, with a total geometrical area of $157 \ \mathrm{cm^2}$. One of two gamma-ray instruments in the ALOFT campaign, iSTORM recorded glows, terrestrial gamma-ray flashes (TGFs), and the newly discovered flickering gamma-ray flashes (FGFs).

\end{abstract}



\begin{keyword}
Terrestrial gamma-ray flash
 \sep Gamma-ray glows \sep Flickering gamma-ray flashes \sep Scintillators \sep Gamma-ray instruments \sep airborne sciences



\end{keyword}

\end{frontmatter}




\section{Introduction}
\label{sec1}

\placetextbox{0.5}{0.05}{\large\textsf{DISTRIBUTION STATEMENT A. Approved for public release: distribution is unlimited.}}%

Two major gamma-ray phenomena associated with thunderstorms are known as terrestrial gamma-ray flashes (TGFs)~\cite{TGF_OG}, which are large bursts of gamma rays that occur in the order of 10--100s of microseconds, and glows~\cite{glows}, which are transients that can last for periods of seconds to minutes. Both phenomena originate from similar mechanisms in that intense electric fields accelerate electrons to relativistic speeds, inducing electron avalanches followed by high-energy photon emission through Bremsstrahlung. Although the phenomena were discovered over 30 years ago, many questions remain, such as the underlying conditions that produce TGFs as well as the temporal and spatial extent of glows.

Space-based instruments, such as BATSE, Fermi, AGILE, RHESSI, and ASIM, have all observed TGFs from space~\cite{TGF_OG, FermiTGF, AGILE, RHESSI, ASIM}. Although they were successful in detecting TGFs, questions remain as to their prevalence~\cite{RareTGFs, SmithPrevelance} and intensity spectrum~\cite{DwyerIntensity}. Space-based instruments have relatively low revisit rates with a rapid fly-by speed, and as a result, were thought to be a rare phenomenon~\cite {IngridPaper}. We note that ground stations have also been able to study gamma-ray emissions from thunderstorms~\cite{Chaffin, DownTGFs}.

Airborne deployments of instruments offer an alternative platform to survey gamma-ray emissions from thunderstorms. However, such deployments prove to be difficult due to the dangers of flying above or around active thunderstorms. A NASA F-106 flew into an active thunderstorm, demonstrating the dangers of flying as such, and also coincidentally discovered the first gamma-ray emissions from thunderstorms, a phenomenon now known as gamma-ray glows~\cite{glows}. Several efforts have been made over the years to deploy instruments on airborne platforms. ADELE deployed on a Gulfstream V~\cite{ADELE} and a WC-130~\cite{ADELE-Hur}. ILDAS flew on the Airbus A340 and observed gamma-ray glows~\cite{IDAL}. The GOES validation campaign flew an ER-2 aircraft over active thunderstorms and observed glows~\cite{GoesGlows}. In addition, a balloon-borne campaign also measured gamma-ray emissions during active thunderstorms~\cite{UAH-Balloon}. Instruments such as XStorm, THOR, and HERA are also in the process of airborne deployments~\cite{XStorm, THOR, HERA}.

The ALOFT campaign~\cite{lang2025hunting} was designed to study these gamma-ray phenomena by flying a suite of instruments spanning multiple wavelengths (FEGS, AMPR, CRS, EXRAD, CoSSIR)~\cite{FEGS, ampr, CRS, exrad, cossir} and sensors to study the electric field (EFCM, LIP)~\cite{er2Glow, LIP}. Two gamma-ray instruments flew onboard, the BGO instrument~\cite{er2Glow}, led by the University of Bergen (UiB), and iSTORM (this work). The UiB instrument was a single module of the BGO high-energy detector on ASIM~\cite{ASIM}. The motivation for flying two independent gamma-ray instruments is to provide cross-validation for any observed phenomenology.

\begin{figure}
  \centering
  \includegraphics[trim={0cm 0cm 0cm 0cm}, clip, width=\linewidth]{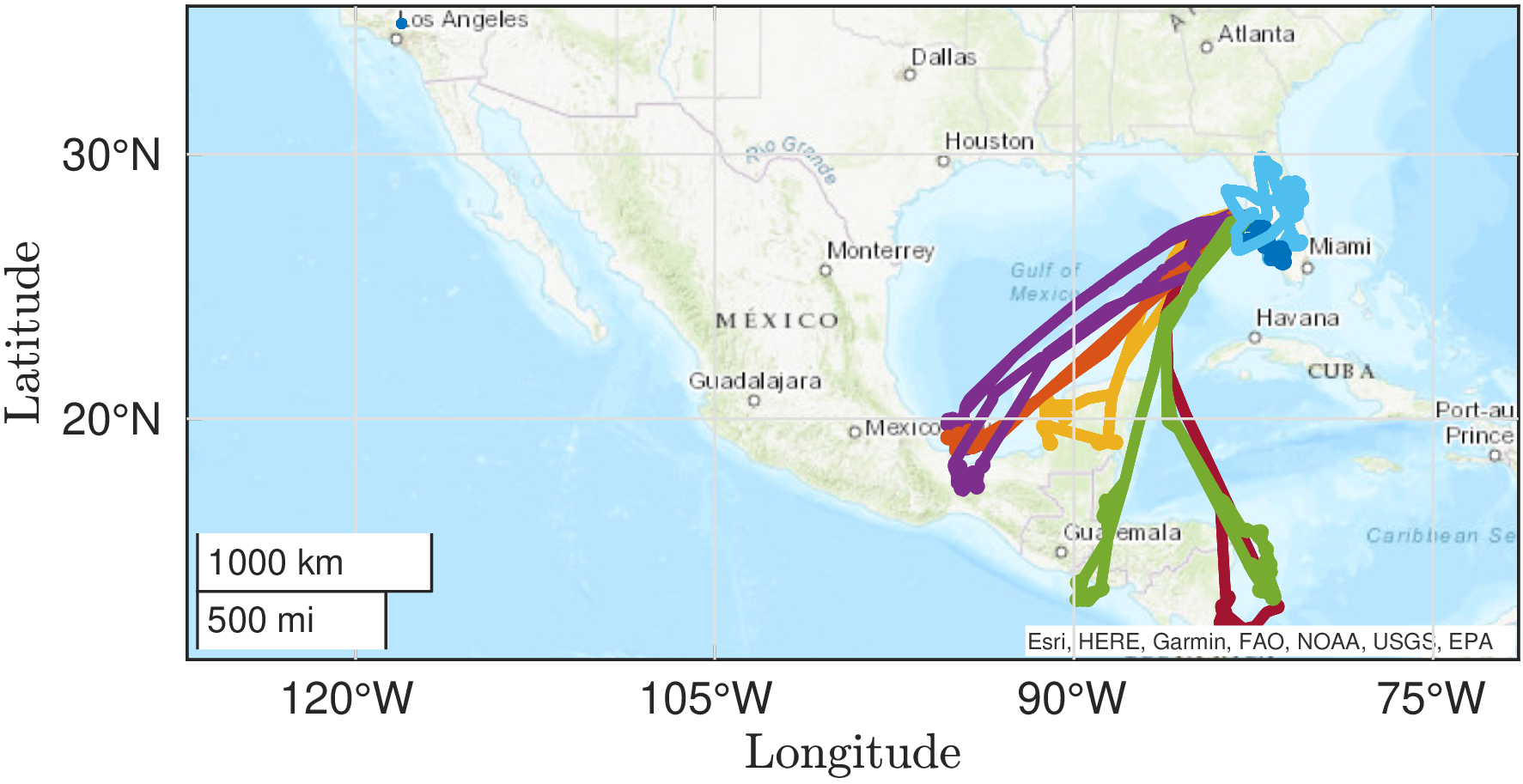}
  \caption{GPS tracks of the 10 science flights during the ALOFT campaign, with each color representing a distinct flight path.}
  \label{fig:aloftTracks}
\end{figure}

The ALOFT campaign was executed on a NASA ER-2~\cite{er2} that flew 10 flights over Central America and the Caribbean, each spanning 3-8 hours. Fig.~\ref{fig:aloftTracks} plots the tracks of the 10 science flights originating from MacDill Air Force Base in Tampa, Florida, the home base of operations.

This manuscript focuses on the development of iSTORM (in-Situ Thunderstorm Observer for Radiation Mechanisms) and highlights some of its results from the ALOFT campaign. Sec.~\ref{sec:iSTORM} discusses the construction and characteristics of iSTORM. Sec.~\ref{sec:results} presents selected results for the flight campaign.

\begin{figure}[h]
  \centering
  \includegraphics[trim={1cm 3cm 2cm 3cm}, clip, width=0.75\linewidth]{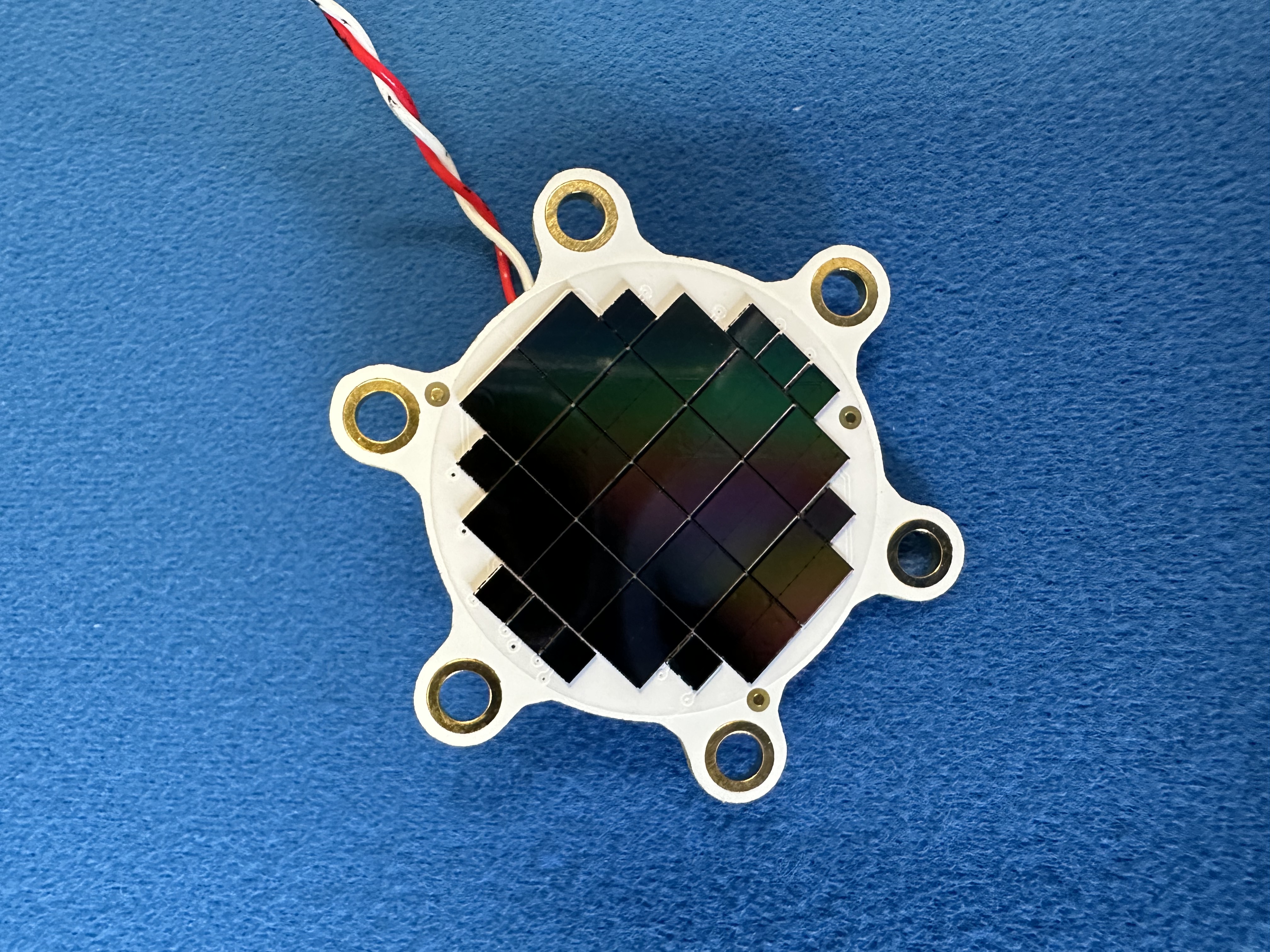}
  \caption{Picture of a custom SiPM array developed for iSTORM. The SiPM array is then coupled to an individual $\mathrm{CeBr}_3$ scintillator.}
  \label{fig:sipm}
\end{figure}

\section{Development of the iSTORM Instrument}
\label{sec:iSTORM}

iSTORM employs 32 $\mathrm{CeBr}_3$ scintillators, each a 1-inch right cylinder with a height of 1 inch. We chose to segment the detectors to reduce pulse pileup. Each scintillator crystal is read by a custom SiPM array and biased to 26V (featured in Fig.~\ref{fig:sipm}). The SiPMs are optically coupled using a silicone optical pad (EJ-560). The average resolution of all the detectors is $4.4 \%$ full-width-at-half-maximum (FWHM) at $662 \ \mathrm{keV}$. We also implement a plastic scintillator and a bare SiPM. The SiPMs are read out by a CAEN A5202~\cite{CAEN-FERS, CAEN_2026}, which is a 64 channel front-end readout system for SiPMs which uses a CITIROC 1A application-specific integrated circuit (ASIC)~\cite{WeeROC}. The ASIC reads out, amplifies, shapes, and digitizes signals while the rest of the front end completed the rest of the digital signal processing via FPGA and communication interfaces~\cite{fers5202}. The front end is configured with a 12-bit ADC and dual-gain stages. The system is set to spectroscopy mode with zero suppression. Note that integration onto the aircraft, 1 crystal disconnected by accident, which was not fixed due to scheduling.

A BeagleBone Black (BBB) serves as the instrument computer, which controls the front end and a GPS receiver (CD PA1616S). The GPS receiver generates a pulse-per-second (PPS), which is then introduced to an ASIC channel and used to discipline the front-end's clock. Data is written to an on-board flash drive. The A5202's data stream contains an event counter. Therefore, if the BBB drops packets in high-count-rate events, the event counter can recover the number of lost events. However, we cannot retrieve the energy or the accurate time of the lost events. We therefore distribute the number of lost events uniformly between two recorded events. The instrument also contains an Arduino that measures temperature throughout the system~\cite{Arduino}, but it was not utilized in flight as it introduced noise to the front end system.  

\begin{figure*}[h]

\begin{center}

\tikzset{every picture/.style={line width=0.75pt}} 

\begin{tikzpicture}[x=0.75pt,y=0.75pt,yscale=-1,xscale=1]

\draw   (20,80) -- (110,80) -- (110,120) -- (20,120) -- cycle ;
\draw   (10,20) -- (120,20) -- (120,60) -- (10,60) -- cycle ;
\draw    (60,60) -- (60,80) ;
\draw   (130,80) -- (210,80) -- (210,120) -- (130,120) -- cycle ;
\draw    (110,100) -- (130,100) ;
\draw   (130,130) -- (210,130) -- (210,170) -- (130,170) -- cycle ;
\draw    (120,150) -- (130,150) ;
\draw    (120,100) -- (120,150) ;
\draw   (230,130) -- (310,130) -- (310,170) -- (230,170) -- cycle ;
\draw    (220,150) -- (230,150) ;
\draw    (210,150) -- (220,150) ;
\draw   (330,130) -- (410,130) -- (410,190) -- (330,190) -- cycle ;
\draw    (310,150) -- (330,150) ;
\draw   (230,80) -- (310,80) -- (310,120) -- (230,120) -- cycle ;
\draw    (210,100) -- (230,100) ;
\draw  [dash pattern={on 0.84pt off 2.51pt}]  (310,100) -- (360,100) ;
\draw  [dash pattern={on 0.84pt off 2.51pt}]  (360,100) -- (360,130) ;
\draw   (240,20) -- (300,20) -- (300,60) -- (240,60) -- cycle ;
\draw  [dash pattern={on 4.5pt off 4.5pt}]  (271.5,60) -- (271.5,80)(268.5,60) -- (268.5,80) ;
\draw  [dash pattern={on 3.75pt off 3pt on 7.5pt off 1.5pt}]  (300,40) -- (380,40) ;
\draw  [dash pattern={on 3.75pt off 3pt on 7.5pt off 1.5pt}]  (380,130) -- (380,40) ;

\draw (65,40) node  [font=\large,color={rgb, 255:red, 0; green, 0; blue, 0 }  ,opacity=1 ] [align=left] {\begin{minipage}[lt]{85pt}\setlength\topsep{0pt}
\begin{center}
{\fontfamily{cmr}\selectfont Aircraft 28V}
\end{center}

\end{minipage}};
\draw (65.5,100) node  [font=\large,color={rgb, 255:red, 0; green, 0; blue, 0 }  ,opacity=1 ] [align=left] {\begin{minipage}[lt]{68.68pt}\setlength\topsep{0pt}
\begin{center}
{\fontfamily{cmr}\selectfont EMI Filter}
\end{center}

\end{minipage}};
\draw (170,100) node  [font=\large,color={rgb, 255:red, 0; green, 0; blue, 0 }  ,opacity=1 ] [align=left] {\begin{minipage}[lt]{54.4pt}\setlength\topsep{0pt}
\begin{center}
{\fontfamily{cmr}\selectfont 28V-5V}
\end{center}

\end{minipage}};
\draw (170,150) node  [font=\large,color={rgb, 255:red, 0; green, 0; blue, 0 }  ,opacity=1 ] [align=left] {\begin{minipage}[lt]{54.4pt}\setlength\topsep{0pt}
\begin{center}
{\fontfamily{cmr}\selectfont 28V-15V}
\end{center}

\end{minipage}};
\draw (273,150) node  [font=\large,color={rgb, 255:red, 0; green, 0; blue, 0 }  ,opacity=1 ] [align=left] {\begin{minipage}[lt]{65pt}\setlength\topsep{0pt}
{\fontfamily{cmr}\selectfont 12V LDO}
\end{minipage}};
\draw (370,162.25) node  [font=\large,color={rgb, 255:red, 0; green, 0; blue, 0 }  ,opacity=1 ] [align=left] {\begin{minipage}[lt]{54.4pt}\setlength\topsep{0pt}
{\fontfamily{cmr}\selectfont CAEN A5202}
\end{minipage}};
\draw (270,100) node  [font=\large,color={rgb, 255:red, 0; green, 0; blue, 0 }  ,opacity=1 ] [align=left] {\begin{minipage}[lt]{54.4pt}\setlength\topsep{0pt}
\begin{center}
{\fontfamily{cmr}\selectfont BBB}
\end{center}

\end{minipage}};
\draw (352,90.5) node  [font=\large,color={rgb, 255:red, 0; green, 0; blue, 0 }  ,opacity=1 ] [align=left] {\begin{minipage}[lt]{54.4pt}\setlength\topsep{0pt}
{\fontfamily{cmr}\selectfont {\fontsize{0.61em}{0.73em}\selectfont Ethernet}}
\end{minipage}};
\draw (301,40) node  [font=\large,color={rgb, 255:red, 0; green, 0; blue, 0 }  ,opacity=1 ] [align=left] {\begin{minipage}[lt]{74.8pt}\setlength\topsep{0pt}
{\fontfamily{cmr}\selectfont GPS}
\end{minipage}};
\draw (360,29.5) node  [font=\large,color={rgb, 255:red, 0; green, 0; blue, 0 }  ,opacity=1 ] [align=left] {\begin{minipage}[lt]{54.4pt}\setlength\topsep{0pt}
{\fontfamily{cmr}\selectfont {\fontsize{0.61em}{0.73em}\selectfont PPS}}
\end{minipage}};

\end{tikzpicture}

\end{center}

\caption{Block diagram of the iSTORM sub-systems}
\label{fig:blockDiagram}
\end{figure*}
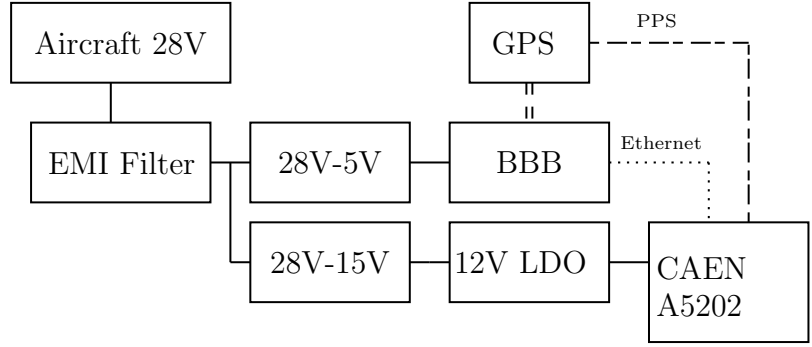

Fig.~\ref{fig:blockDiagram} displays a system level block diagram of iSTORM. The aircraft's 28V supply provides power, which is then filtered with an Electromagnetic Interference (EMI) filter and is converted into 15V and 5V, with the 15V further converted to 12 volt using a low-dropout regulator (LDO) to supply stable power to the front end. The 5V output powers the Beaglebone Black (BBB). 

\begin{figure}[H]
  \centering
  \includegraphics[trim={1cm 2cm 2cm 2cm}, clip, width=0.8\linewidth]{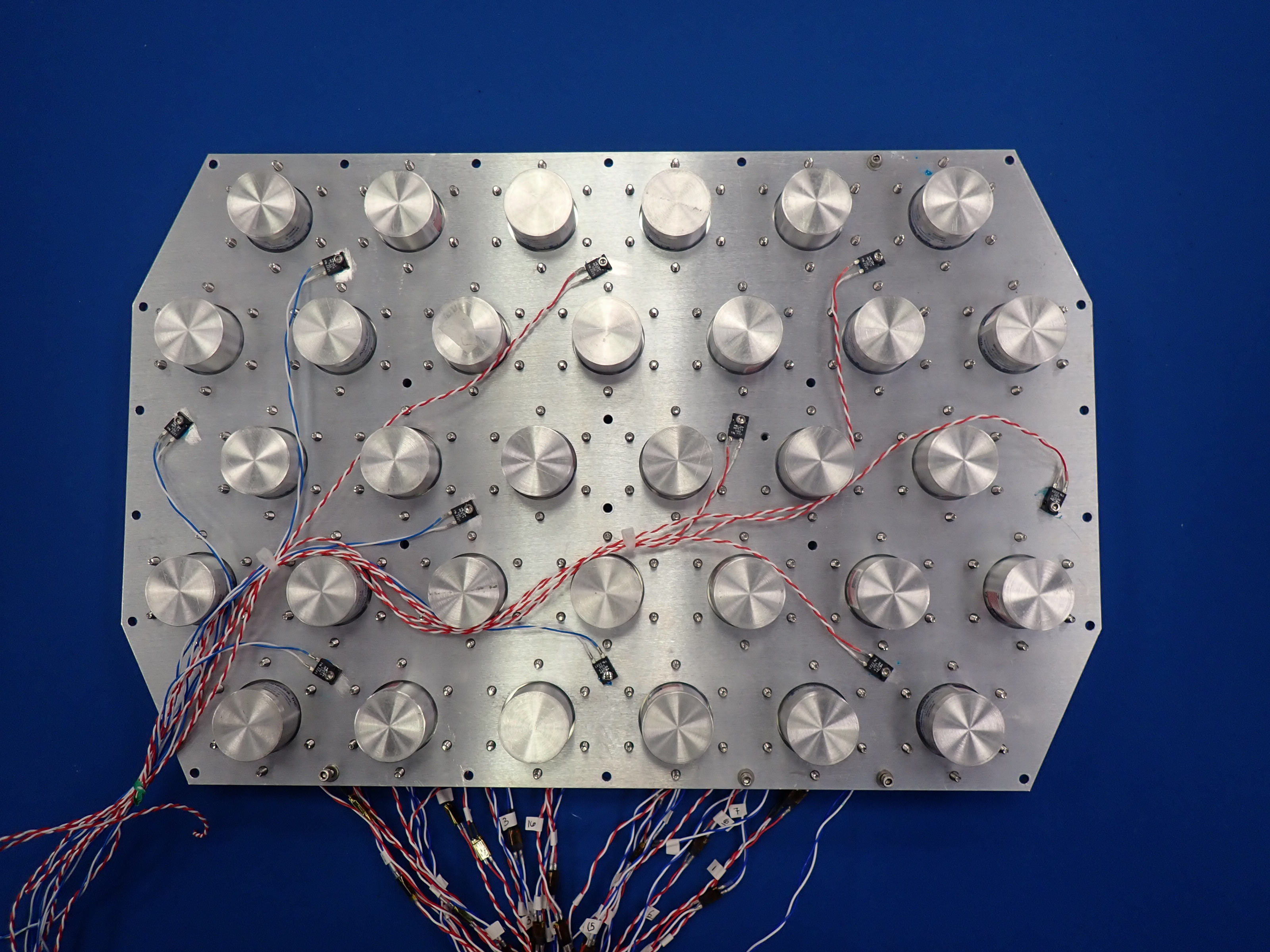}
  \caption{iSTORM's $\mathrm{CeBr}_3$ scintillator array. The wires on the anterior side are the resistive heaters while the posterior side contains the SiPM and associated electronics.}
  \label{fig:array}
\end{figure}

Fig.~\ref{fig:array} shows the $\mathrm{CeBr}_3$ array. Also featured are resistivity heaters. However, they were not utilized in flight as the AC-DC power converter introduced noise into the system.  Once power is provided to the instrument, iSTORM automatically powers up into science mode.

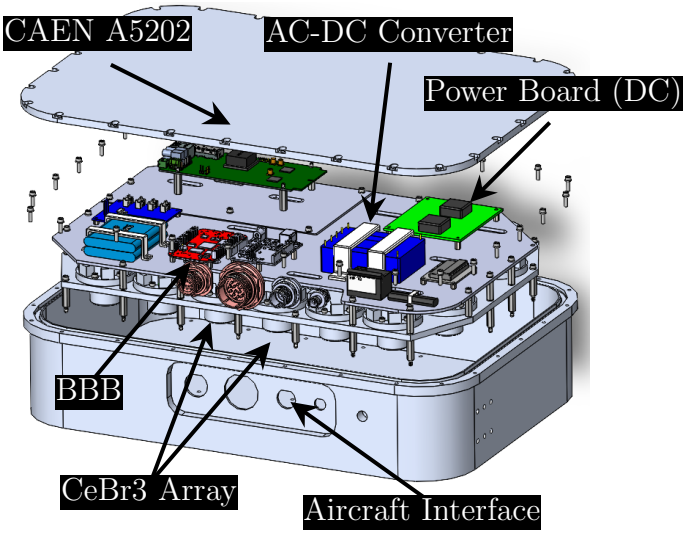
\begin{figure}[h]

\begin{center}

\tikzset{every picture/.style={line width=0.75pt}} 

\begin{tikzpicture}[x=0.75pt,y=0.75pt,yscale=-1,xscale=1]

\draw (145,120.62) node  {\includegraphics[width=217.5pt,height=180.93pt]{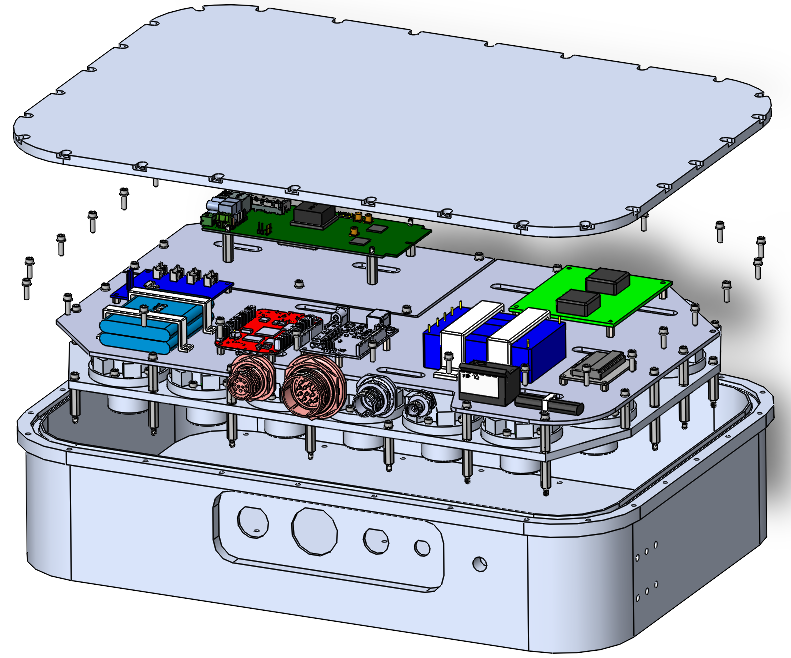}};
\draw [color={rgb, 255:red, 0; green, 0; blue, 0 }  ,draw opacity=1 ][line width=1.5]    (40,190) -- (87.44,133.07) ;
\draw [shift={(90,130)}, rotate = 129.81] [fill={rgb, 255:red, 0; green, 0; blue, 0 }  ,fill opacity=1 ][line width=0.08]  [draw opacity=0] (13.4,-6.43) -- (0,0) -- (13.4,6.44) -- (8.9,0) -- cycle    ;
\draw [color={rgb, 255:red, 0; green, 0; blue, 0 }  ,draw opacity=1 ][line width=1.5]    (210,250) -- (143.25,202.32) ;
\draw [shift={(140,200)}, rotate = 35.54] [fill={rgb, 255:red, 0; green, 0; blue, 0 }  ,fill opacity=1 ][line width=0.08]  [draw opacity=0] (13.4,-6.43) -- (0,0) -- (13.4,6.44) -- (8.9,0) -- cycle    ;
\draw [color={rgb, 255:red, 0; green, 0; blue, 0 }  ,draw opacity=1 ][line width=1.5]    (50,31) -- (106.42,59.21) ;
\draw [shift={(110,61)}, rotate = 206.57] [fill={rgb, 255:red, 0; green, 0; blue, 0 }  ,fill opacity=1 ][line width=0.08]  [draw opacity=0] (13.4,-6.43) -- (0,0) -- (13.4,6.44) -- (8.9,0) -- cycle    ;
\draw [color={rgb, 255:red, 0; green, 0; blue, 0 }  ,draw opacity=1 ][line width=1.5]    (270,60) -- (232.83,97.17) ;
\draw [shift={(230,100)}, rotate = 315] [fill={rgb, 255:red, 0; green, 0; blue, 0 }  ,fill opacity=1 ][line width=0.08]  [draw opacity=0] (13.4,-6.43) -- (0,0) -- (13.4,6.44) -- (8.9,0) -- cycle    ;
\draw [color={rgb, 255:red, 0; green, 0; blue, 0 }  ,draw opacity=1 ][line width=1.5]    (190,30) -- (180.5,106.03) ;
\draw [shift={(180,110)}, rotate = 277.13] [fill={rgb, 255:red, 0; green, 0; blue, 0 }  ,fill opacity=1 ][line width=0.08]  [draw opacity=0] (13.4,-6.43) -- (0,0) -- (13.4,6.44) -- (8.9,0) -- cycle    ;
\draw [color={rgb, 255:red, 0; green, 0; blue, 0 }  ,draw opacity=1 ][line width=1.5]    (70,240) -- (98.6,163.75) ;
\draw [shift={(100,160)}, rotate = 110.56] [fill={rgb, 255:red, 0; green, 0; blue, 0 }  ,fill opacity=1 ][line width=0.08]  [draw opacity=0] (13.4,-6.43) -- (0,0) -- (13.4,6.44) -- (8.9,0) -- cycle    ;
\draw [color={rgb, 255:red, 0; green, 0; blue, 0 }  ,draw opacity=1 ][line width=1.5]    (70,240) -- (127.4,173.04) ;
\draw [shift={(130,170)}, rotate = 130.6] [fill={rgb, 255:red, 0; green, 0; blue, 0 }  ,fill opacity=1 ][line width=0.08]  [draw opacity=0] (13.4,-6.43) -- (0,0) -- (13.4,6.44) -- (8.9,0) -- cycle    ;

\draw (45,195.5) node  [font=\normalsize,color={rgb, 255:red, 248; green, 231; blue, 28 }  ,opacity=1 ] [align=left] {\begin{minipage}[lt]{34pt}\setlength\topsep{0pt}
{\fontfamily{cmr}\selectfont \textcolor[rgb]{1,1,1}{\hl{BBB}}}
\end{minipage}};
\draw (210,255.5) node  [font=\normalsize,color={rgb, 255:red, 248; green, 231; blue, 28 }  ,opacity=1 ] [align=left] {\begin{minipage}[lt]{95.2pt}\setlength\topsep{0pt}
{\fontfamily{cmr}\selectfont \textcolor[rgb]{1,1,1}{\hl{Aircraft Interface}}}
\end{minipage}};
\draw (60,15.5) node  [font=\normalsize,color={rgb, 255:red, 248; green, 231; blue, 28 }  ,opacity=1 ] [align=left] {\begin{minipage}[lt]{95.2pt}\setlength\topsep{0pt}
{\fontfamily{cmr}\selectfont \textcolor[rgb]{1,1,1}{\hl{CAEN A5202}}}
\end{minipage}};
\draw (75,245.5) node  [font=\normalsize,color={rgb, 255:red, 248; green, 231; blue, 28 }  ,opacity=1 ] [align=left] {\begin{minipage}[lt]{74.8pt}\setlength\topsep{0pt}
{\fontfamily{cmr}\selectfont \textcolor[rgb]{1,1,1}{\hl{CeBr3 Array}}}
\end{minipage}};
\draw (280,45.5) node  [font=\normalsize,color={rgb, 255:red, 248; green, 231; blue, 28 }  ,opacity=1 ] [align=left] {\begin{minipage}[lt]{108.8pt}\setlength\topsep{0pt}
{\fontfamily{cmr}\selectfont \textcolor[rgb]{1,1,1}{\hl{Power Board (DC)}}}
\end{minipage}};
\draw (200,16) node  [font=\normalsize,color={rgb, 255:red, 248; green, 231; blue, 28 }  ,opacity=1 ] [align=left] {\begin{minipage}[lt]{108.8pt}\setlength\topsep{0pt}
{\fontfamily{cmr}\selectfont \textcolor[rgb]{1,1,1}{\hl{AC-DC Converter}}}
\end{minipage}};

\end{tikzpicture}

\end{center}

\caption{Exploded view of the iSTORM instrument}
\label{fig:istormExploded}
\end{figure}

We place the detectors and associated electronics inside a hermetically sealed box that maintains 1 atm during flight. Before closure, we purged the instrument with nitrogen gas. The outer envelope is 20 (L) $\times$ 13.75 (W) $\times$ 4.4375 (H) $\mathrm{in}^3$ and the total instrument has a mass of 24.3 kg. The power draw is 0.49 Amps at 28V during science mode. Fig.~\ref{fig:istormExploded} shows an exploded view of iSTORM. The detector array is placed at the bottom level. Right above it sits the associated electronics. The instrument interface includes a 28V input from the aircraft, a debug port, and a connector for the GPS antenna. Due to the short development timeline, real-time telemetry was not enabled for iSTORM.

\subsection{Environmental Testing of iSTORM}
\label{sec:testing}

\begin{figure}[h]
  \centering
  \includegraphics[trim={0cm 0cm 0cm 0cm}, clip, width=1\linewidth]{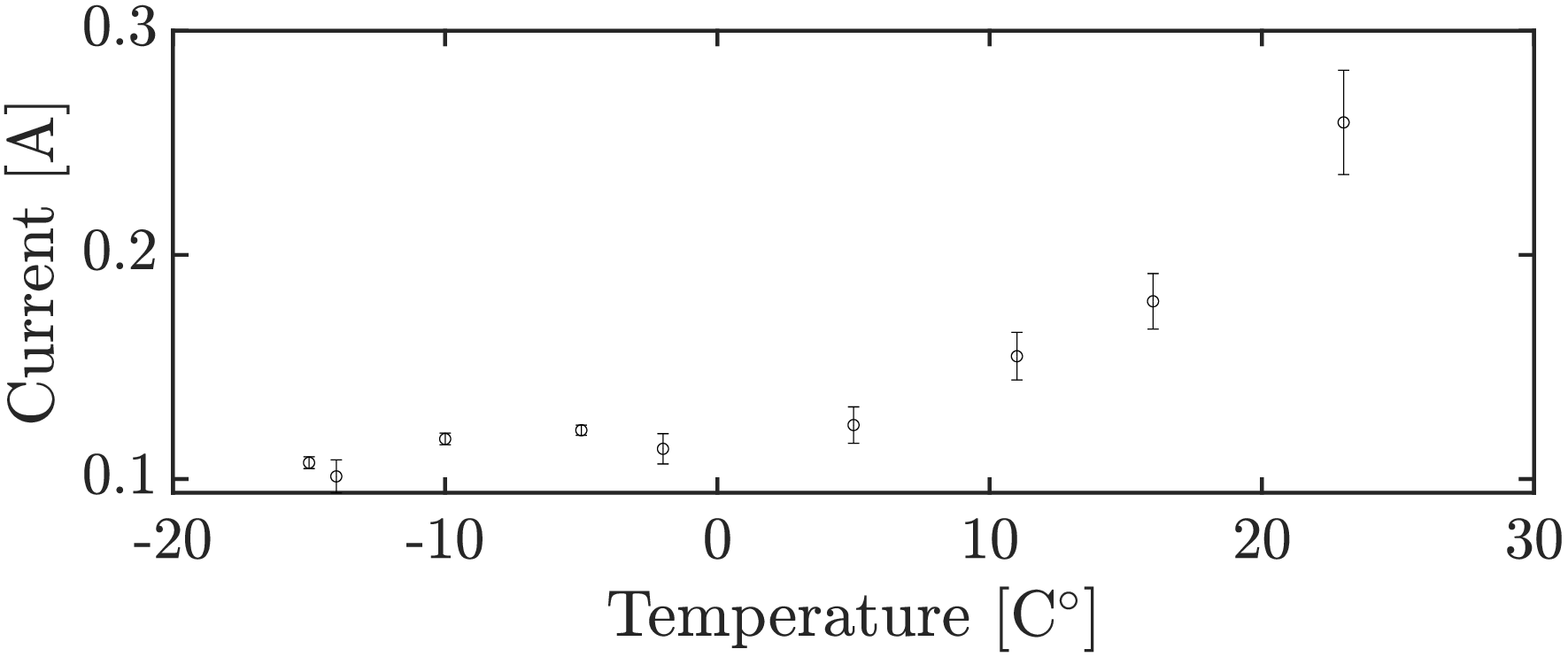}
  \caption{Current draw as a function of ambient temperature required to bias the SiPMs.}
  \label{fig:currentTemp}
\end{figure}

\begin{figure}[h]
  \centering
  \includegraphics[trim={0cm 0cm 0cm 0cm}, clip, width=1\linewidth]{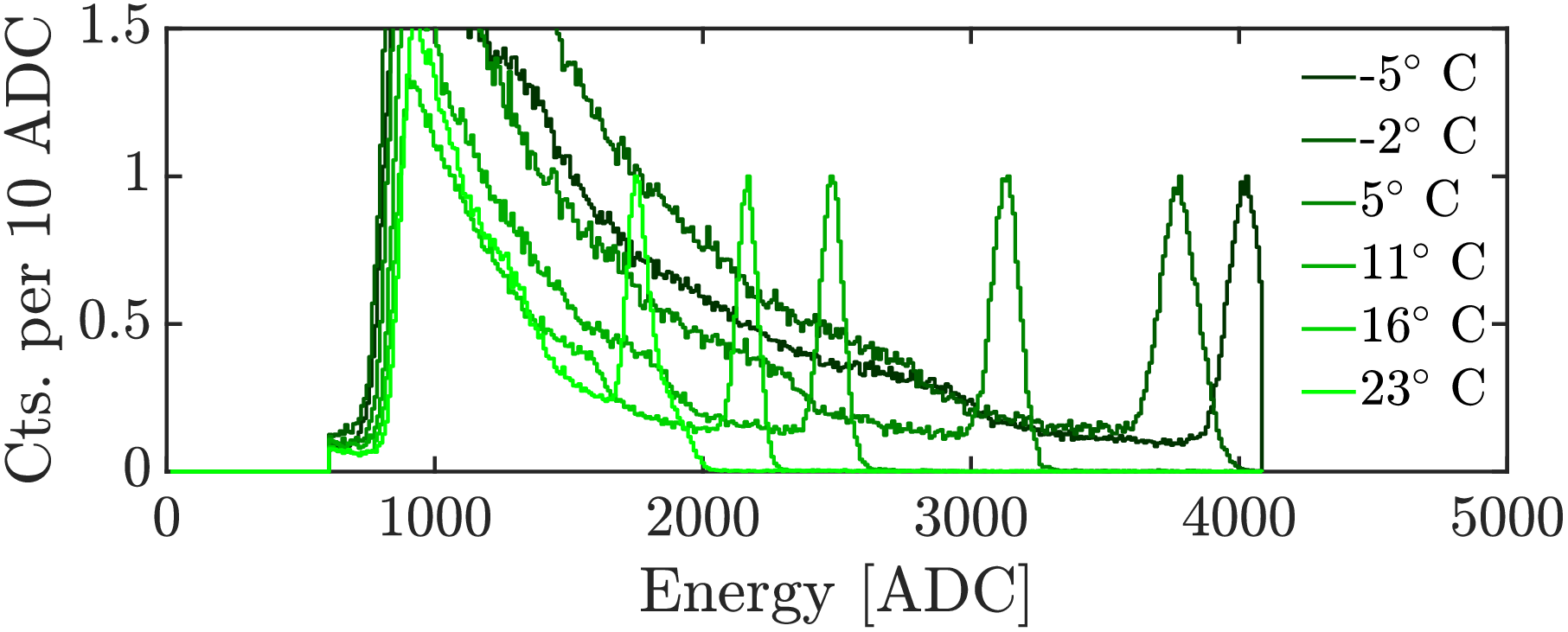}
  \caption{Cs-137 spectrum as a function of different temperatures. Counts are normalized to the Cs-137 photopeak.}
  \label{fig:spectraTemp}
\end{figure}

We conducted several environmental tests for instrument commissioning. First, iSTORM was tested in a low-pressure environment equivalent to an altitude of $90 \ \mathrm{kft}$. In addition, rapid chamber evacuation and pressurization were also tested to stress the hermetic seals and evaluate the safety of the pressure vessel. Next, the system underwent thermal cycling at temperatures ranging from $-15$ to $23 \ ^{\circ}\mathrm{C}$. Higher temperature tests were not conducted, as the instrument's operational environment at altitude is predominantly cold. Fig.~\ref{fig:currentTemp} plots the current draw required to bias the SiPMs over the environmental chamber's temperature settings. As expected, the current draw increases as a function of temperature. Fig.~\ref{fig:spectraTemp} plots Cs-137 spectra taken at different temperatures to characterize the spectral response to temperature variations. The x-axis represents analog to digital converter (ADC) units. As expected, the photopeak shifts to higher ADC values with decreasing temperature. A major contributor to this is the SiPM's gain, which is temperature dependent, and the scintillator's temperature-dependent light yield~\cite{cebr3_temp}. The gain of the SiPM increases enough such that at $-5^{\circ}\mathrm{C}$, the photopeak is cropped into the overflow bin. As we wish to maintain an energy range that can encompass the $511 \ \mathrm{keV}$ peak, we define $-5^{\circ}\mathrm{C}$ as a lower limit for science operations.

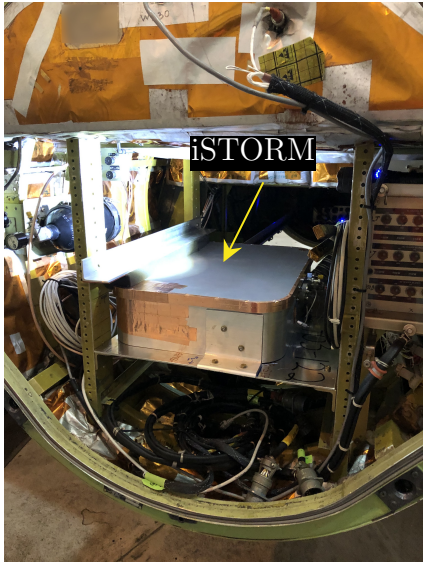
\begin{figure}[h]

\begin{center}

\tikzset{every picture/.style={line width=0.75pt}} 

\begin{tikzpicture}[x=0.75pt,y=0.75pt,yscale=-1,xscale=1]

\draw (105,140) node  {\includegraphics[angle=0,width=157.5pt,height=210pt]{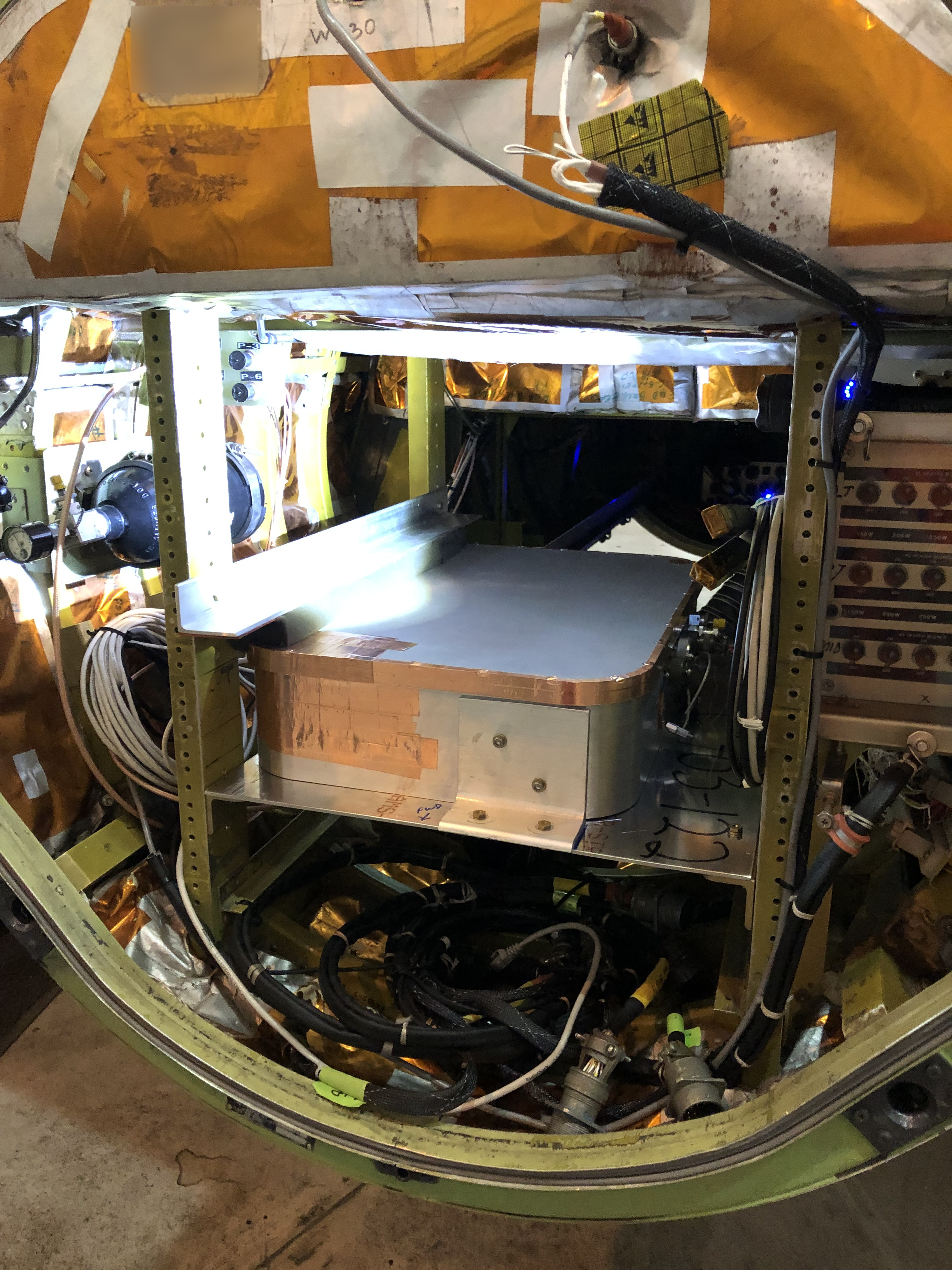}};
\draw [color={rgb, 255:red, 248; green, 231; blue, 28 }  ,draw opacity=1 ]   (130,90) -- (111.34,127.32) ;
\draw [shift={(110,130)}, rotate = 296.57] [fill={rgb, 255:red, 248; green, 231; blue, 28 }  ,fill opacity=1 ][line width=0.08]  [draw opacity=0] (10.72,-5.15) -- (0,0) -- (10.72,5.15) -- (7.12,0) -- cycle    ;

\draw (130,75.5) node  [font=\large,color={rgb, 255:red, 248; green, 231; blue, 28 }  ,opacity=1 ] [align=left] {\begin{minipage}[lt]{60pt}\setlength\topsep{0pt}
{\fontfamily{cmr}\selectfont \textcolor[rgb]{1,1,1}{\hl{iSTORM}}}
\end{minipage}};

\end{tikzpicture}
\end{center}

\caption{The iSTORM instrument mounted in the ER-2 right wing pod.}
\label{fig:istormInPodImg}
\end{figure}

\subsection{iSTORM and ALOFT Flight Campaign}
\label{sec:iSTORMinAloft}

iSTORM was mounted in the midbody of the ER-2's right-wing pod. Fig.~\ref{fig:istormInPodImg} shows the instrument mounted. Fastened onto an aluminum plate in the center of the pod, it sits right under the wing. The pod is fitted with blowers that introduce hot air during flight to maintain non-freezing stable temperatures. The pod is closed with the pod's nose cone, which houses FEGS and the BGO instrument.

During science flights, the aircraft would normally cruise at altitudes between 65-68 kft (19.8-20.7 km) at a cruising speed of around 450 miles per hour (200 meters per second). During the science portion of the flight, the pilot would be instructed to fly over active regions of the thunderstorm. Since real-time data is telemetered to the ground from the instruments, the gamma-ray count rate could be observed from the BGO instrument. Should we observe an increase in the count rate, the pilot would then be instructed to return to the point of interest and bow-tie over it while maintaining wings level. However, during certain periods, the pilot took discretion due to their rather visually accessible and up-to-date weather conditions, and the fact that certain thunderstorms were so active that no waypoints were required.

\section{Observed Phenomenology in Flight}
\label{sec:results}

This section reports on some results observed during the flight on July 24, 2023. The flight focused on a thunderstorm region right off the coast of Veracruz, Mexico, in the Bay of Campeche. This flight encountered the campaign's most active thunderstorm system, during which nearly 100 TGFs and numerous gamma-ray glows were observed. Also observed is a newly discovered phenomenon called Flickering Gamma-ray Flash (FGF). The BGO's observation is consistent with that of iSTORM.

\begin{figure}[H]
  \centering
  \includegraphics[trim={0cm 0cm 0cm 0cm}, clip, width=1\linewidth]{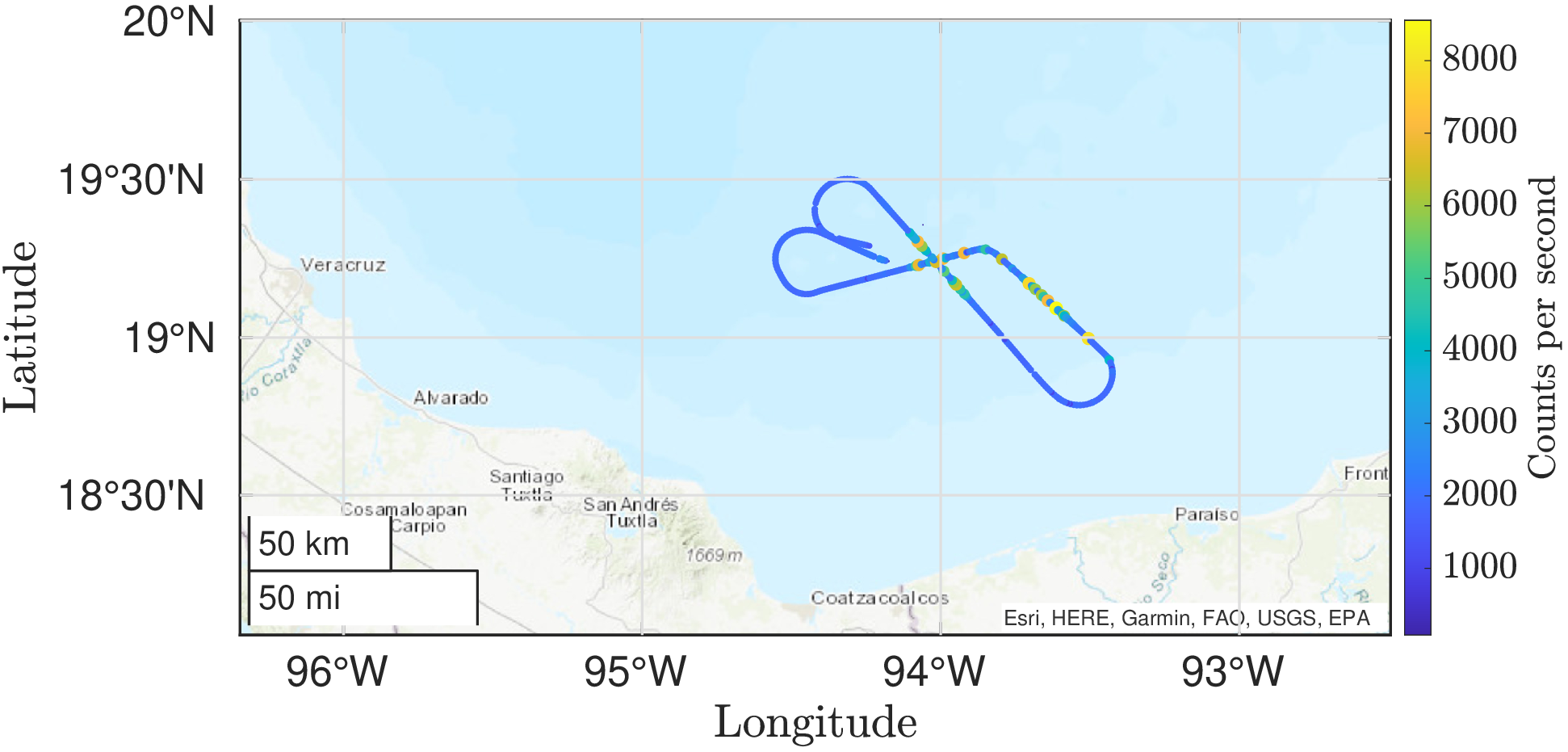}
  \caption{GPS tracks of a segment of the July 24, 2023 flight. The marker color and size correspond to the count rate of that location. Each point represents a 5-second bin.}
  \label{fig:glowCtsGPS}
\end{figure}

\begin{figure}[H]
  \centering
  \includegraphics[trim={0cm 0cm 0cm 0cm}, clip, width=1\linewidth]{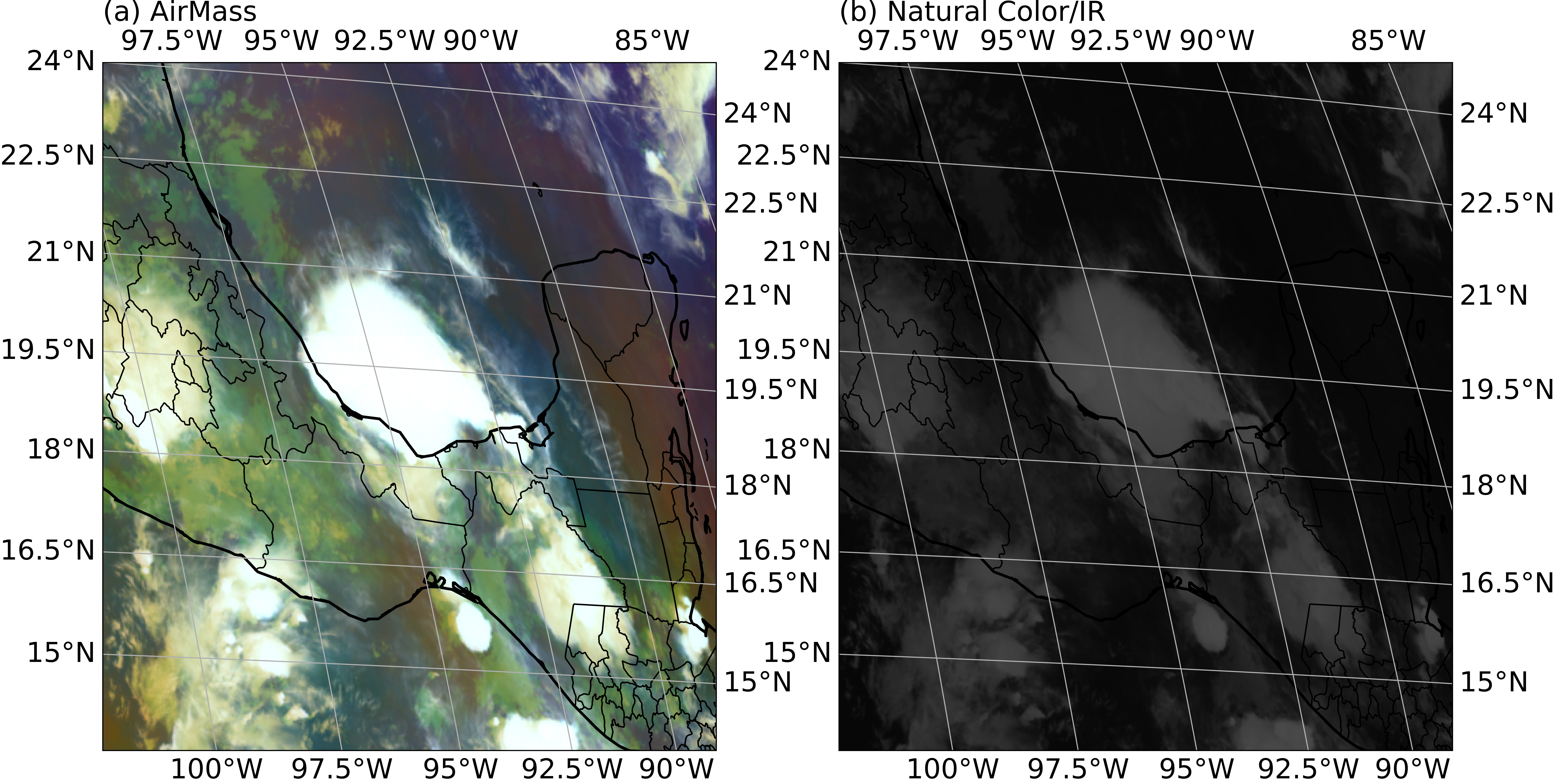}
  \caption{Geostationary Operational Environmental Satellite
 (GOES) Advanced Baseline Imager (ABI) view of the flight trajectory shown in Fig.~\ref{fig:glowCtsGPS}. (a) Plots the AirMass view while (b) plots the Natural Color and Infrared mix.}
  \label{fig:goesABI}
\end{figure}

Fig.~\ref{fig:glowCtsGPS} plots a 30-minute segment of that flight and features a time series that contains glows, TGFs, and FGFs. The segment demonstrates a bow-tie maneuver over an active gamma-ray cloud. Fig.~\ref{fig:goesABI} shows the Geostationary Operational Environmental Satellite's
 (GOES) Advanced Baseline Imager (ABI) view of the flight segment mentioned earlier, and aims to give meteorological context~\cite{GOES_ABI, GOES-Data}. Fig.~\ref{fig:goesABI}a shows the AirMass view, which is a mixture of the different red, green, and blue channels that highlights environmental characteristics~\cite{GoesAirMass}. The white system, shown in the center of the figure, over the southern area of the Gulf of Mexico/Bay of Campeche denotes a very high thick cloud top. They are associated with deep convection, featuring strong moisture and upper-level clouds, and are likely cumulonimbus systems. Fig.~\ref{fig:goesABI}b shows the natural color/infrared (IR). As the flight was at night, the view was dominated by the infrared channel.

\subsection{Gamma-ray Glows}
\label{sec:glows}

Gamma-ray glows are enhancements in the gamma-ray background first reported by Parks et al.~\cite{glows} that could last for seconds or more. They have since been observed by balloon-borne instruments~\cite{glowBalloon}, from the ground~\cite{japanGroundGlow}, and the ER-2 platform~\cite{er2Glow}. An extensive review of glows observed by ALOFT is available in~\cite{MartinoGlow}.

\begin{figure}[H]
  \centering
  \includegraphics[trim={0cm 0cm 0cm 0cm}, clip, width=1\linewidth]{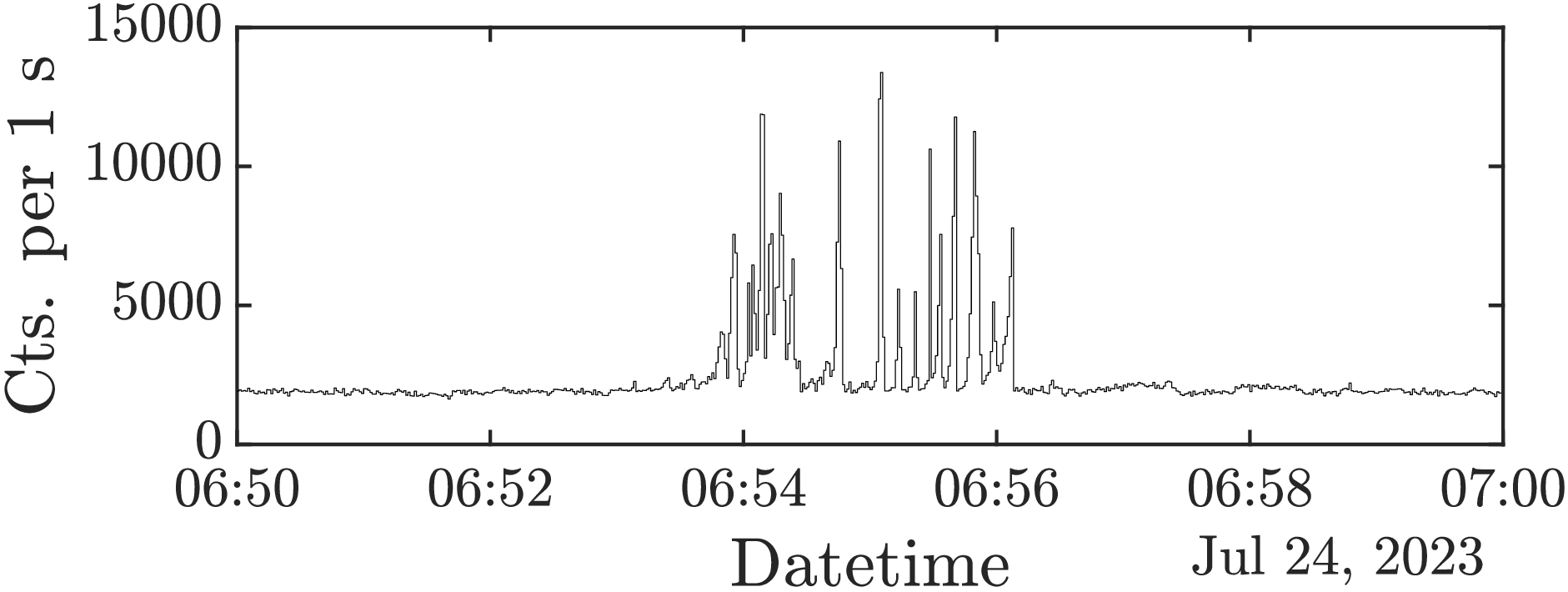}
  \caption{Time series of a pass over an active thunderstorm region. Time is in UTC.}
  \label{fig:glowCts}
\end{figure}

Fig.~\ref{fig:glowCts} plots the gamma count rate over 10-minute period. The flyover reveals the region to be very active with gamma-ray emissions for almost 2.5 minutes, during which the aircraft traveled around 30 km (18.6 miles). In the series, the longest glow lasts almost 9 seconds. Moreover, their temporal variability is structured, likely indicating they are complex phenomena and are unstable electric fields.

\begin{figure}[H]
  \centering
  \includegraphics[trim={0cm 0cm 0cm 0cm}, clip, width=1\linewidth]{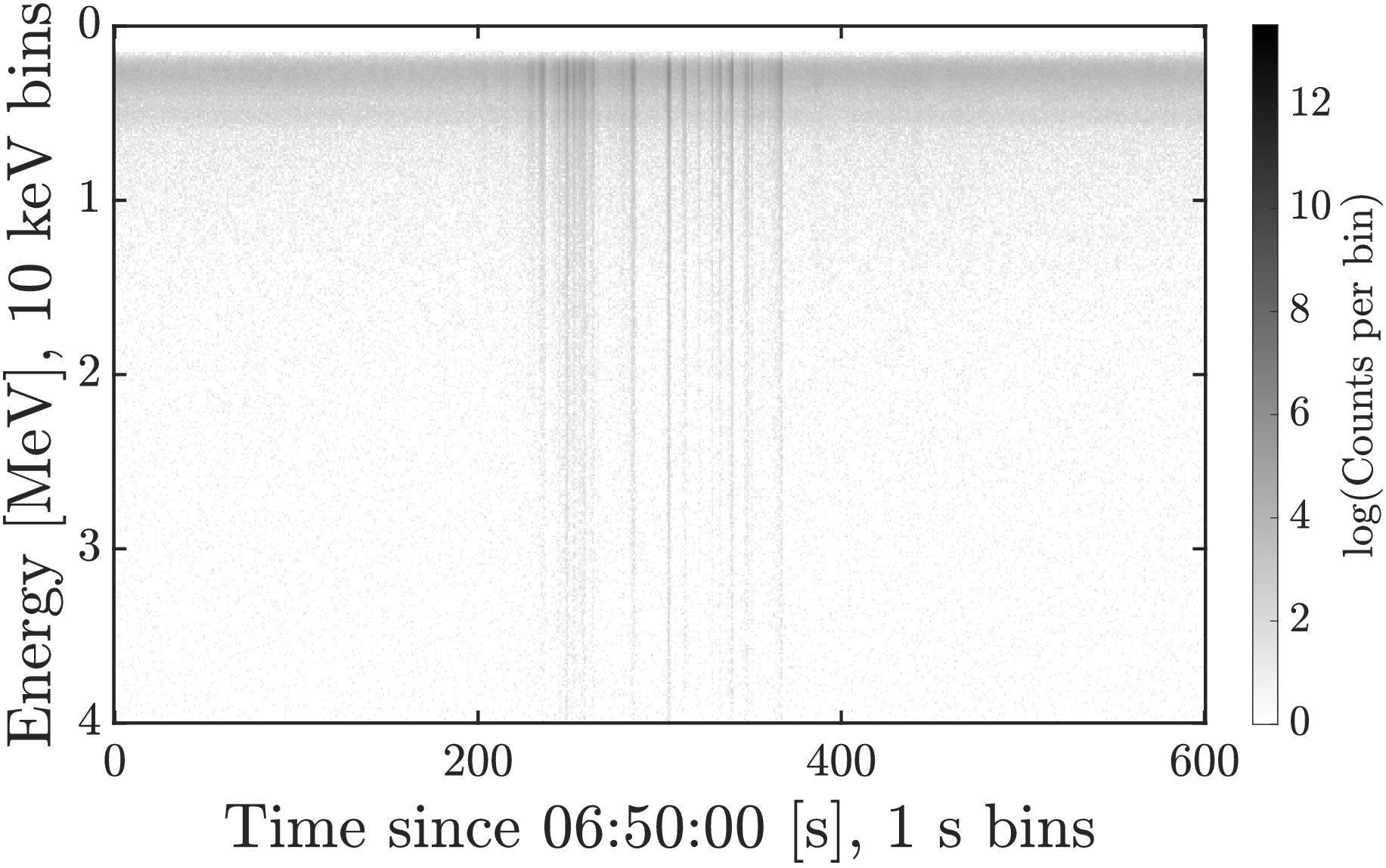}
  \caption{Recorded temporal-energy response of the 10-minute case study. Time is in UTC.}
  \label{fig:glowSpectrum}
\end{figure}

Fig.~\ref{fig:glowSpectrum} further explores the glows by plotting the energy-time relationship. When over non-active regions, a 511 keV background is observed seen in the horizontal band at 511 keV present at all times. During glows, we observe an increase in flux with energies well above 4 MeV.

\subsubsection{Energy Spectra Observed During Gamma-ray Glows}
\label{sec:glowSpectrum}

\begin{figure}[h]
  \centering
  \includegraphics[trim={0cm 0cm 0cm 0cm}, clip, width=1\linewidth]{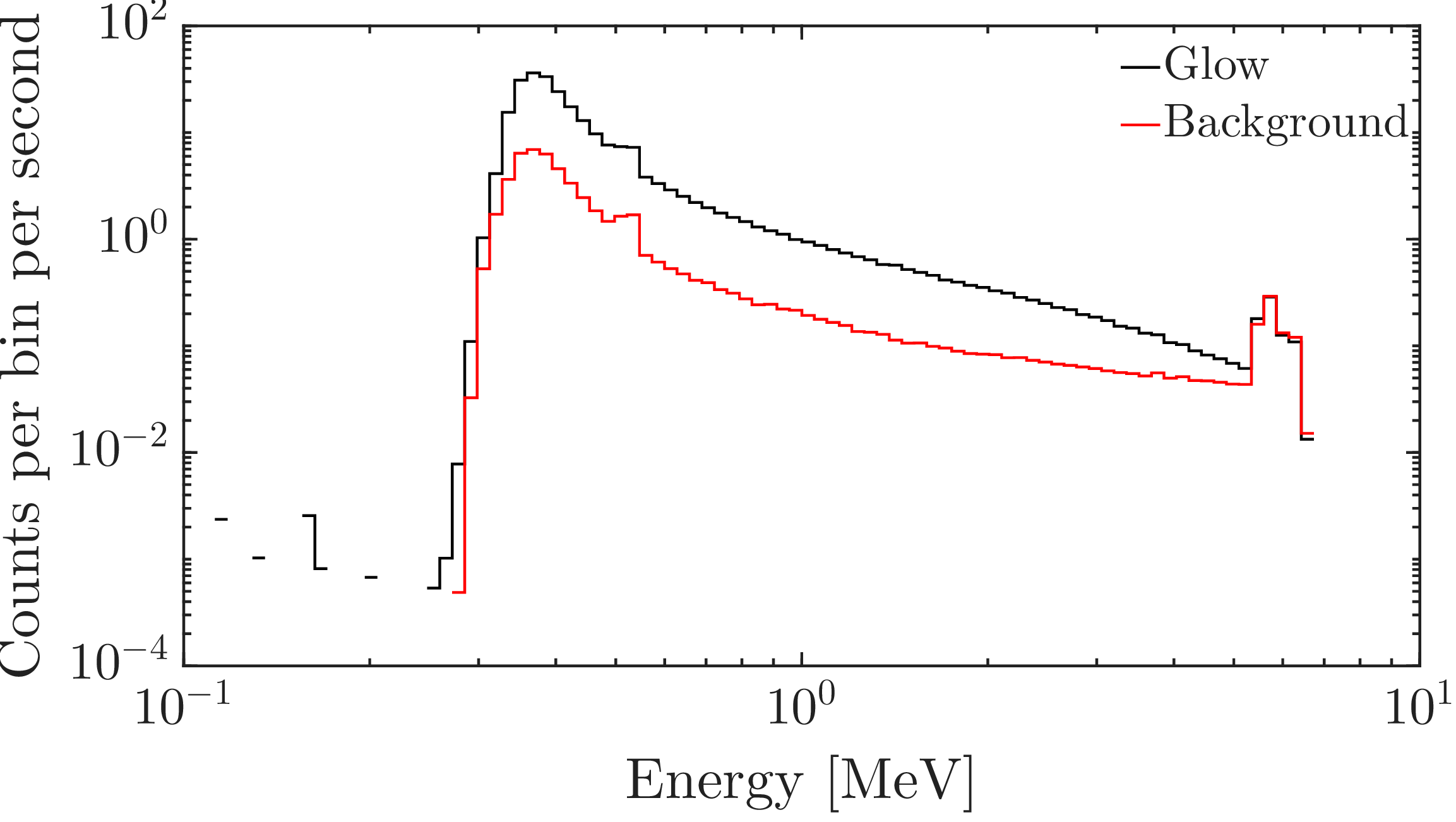}
  \caption{Comparison of the measured spectrum during active glows vs background. }
  \label{fig:glowErg}
\end{figure}

Fig.~\ref{fig:glowErg} plots the energy spectra observed by iSTORM during an active glow vs the surrounding background. The glow curve represents 162 seconds of averaged spectrum when the count rate is at least twice the background. The background represents the same acquisition time, but for a time frame right before the glow episode. We can clearly observe the 511 keV mostly originating from electron-positron annihilation triggered by high-energy cosmic rays interacting with Earth's atmosphere~\cite{REGENER1934}. The large peak spectral feature above 5 MeV represents the overflow bins for each detector channel.

The spectra have not yet been deconvolved to remove the effects of the instrument response and the surrounding environment (e.g., the aircraft and atmosphere). Therefore, no real implications can be deduced as to the spectral shape of the source. However, work is underway to deconvolve those responses to back out the source spectrum and other characteristics~\cite{DavidEnergyPaper}.

\subsection{Terrestrial Gamma-ray Flashes}
\label{sec:tgf}

\begin{figure}[H]
  \centering
  \includegraphics[trim={0cm 0cm 0cm 0cm}, clip, width=1\linewidth]{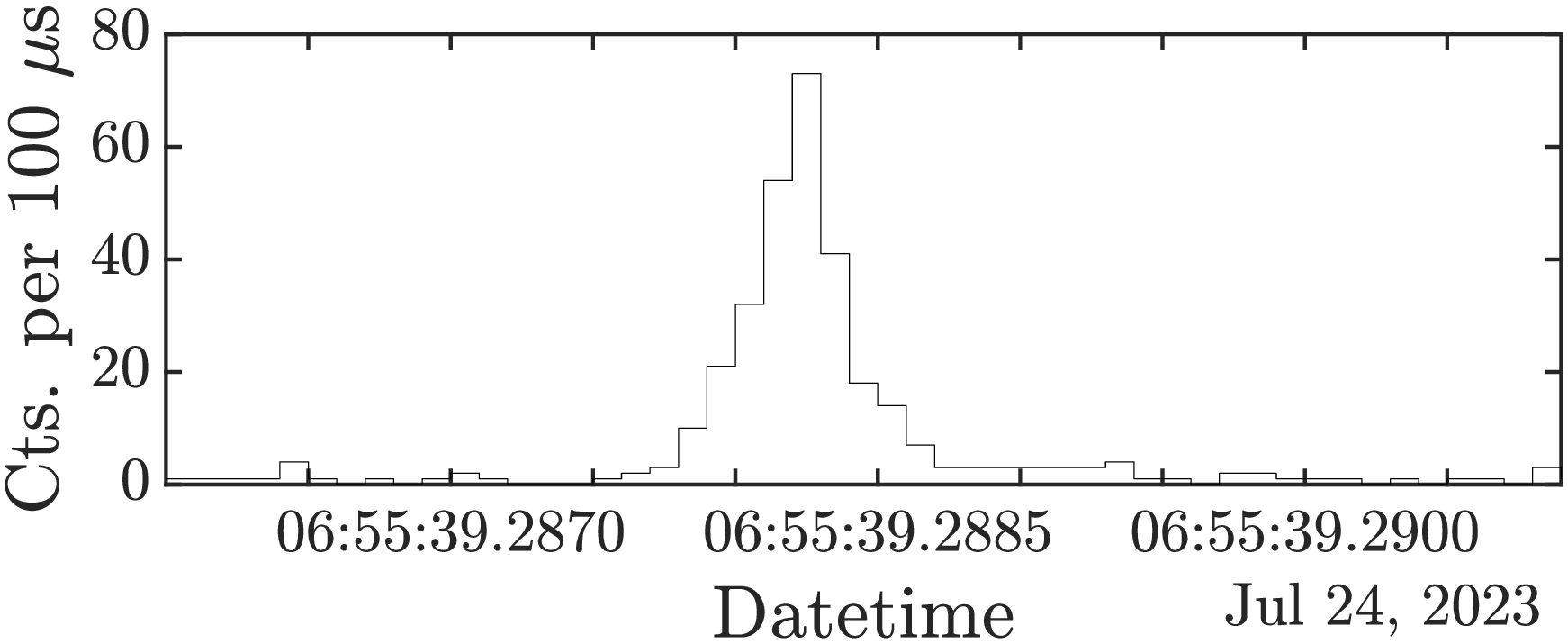}
  \caption{Example of a TGF detected by iSTORM that lasts almost $1 \ \mathrm{ms}$. Time is in UTC.}
  \label{fig:tgf2}
\end{figure}

Terrestrial gamma-ray flashes (TGFs) were first reported in 1994 with detections from the Burst and Transient Source Experiment (BATSE) on the Compton Gamma Ray Observatory (CGRO)~\cite{TGF_OG}. They are characterized by a short burst of gamma rays lasting $10 \ \mu \mathrm{s}$ to $1 \ \mathrm{ms}$. To date, the majority of observations were done by orbiting satellites with a relatively low number of airborne and ground measurements~\cite{TGF_rarity, TGF_Ground, TGF_japan}. ALOFT has detected over 100 TGFs over its entire campaign. In the 10-minute range we present in Fig.~\ref{fig:glowCts}, iSTORM detected 7 TGFs, which are buried in the glow episodes and therefore not clearly visible in this 1 Hz binning. Fig.~\ref{fig:tgf2} presents an example of a TGF detected by iSTORM and lasts almost $1 \ \mathrm{ms}$. Overall, the ALOFT campaign discovered a whole new distribution of TGFs, of which evidence has been found that a weak population of TGFs exists that cannot be observed from space~\cite{IngridPaper, FuglestadPaper}.

\subsection{Flickering Gamma-ray Flashes}
\label{sec:fgf}

\begin{figure}[H]
  \centering
  \includegraphics[trim={0cm 0cm 0cm 0cm}, clip, width=1\linewidth]{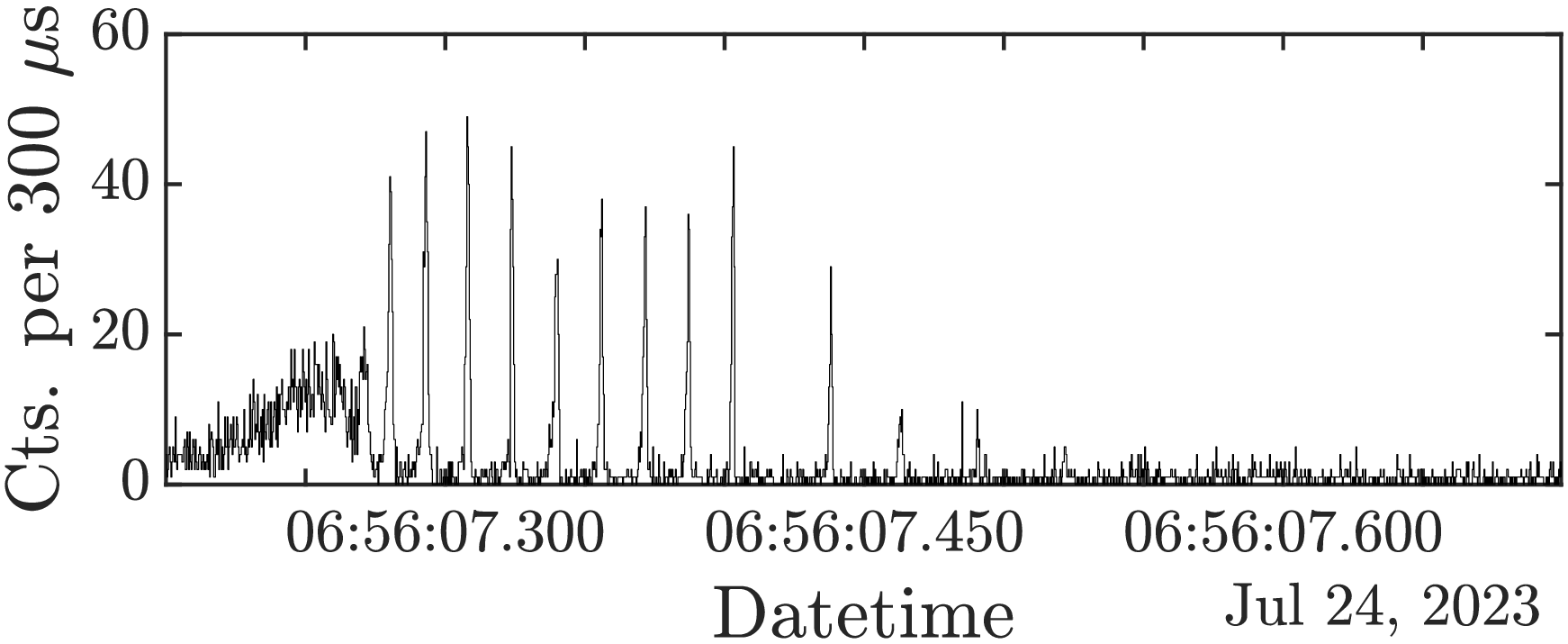}
  \caption{Example of an FGF detected by iSTORM. Time is in UTC.}
  \label{fig:fgf}
\end{figure}

FGFs are a newly observed phenomenon first observed in the ALOFT campaign. FGFs are marked by multiple pulses that we observed to span between 20 and 250 ms. Each pulse is observed to have a duration longer than an ordinary TGF. The details of the observations of FGFs are reported in~\cite{FGF}. In the 10-minute case study presented in this section, 3 FGFs were detected by iSTORM. Fig.~\ref{fig:fgf} plots an FGF that endures around 250 ms with around 14 pulses detected. The first 10 pulses are, on average, spaced 15 ms apart with each pulse lasting almost 3 ms. Over the entire campaign, 24 FGFs were observed.

\begin{figure}[H]
  \centering
  \includegraphics[trim={0cm 0cm 0cm 0cm}, clip, width=1\linewidth]{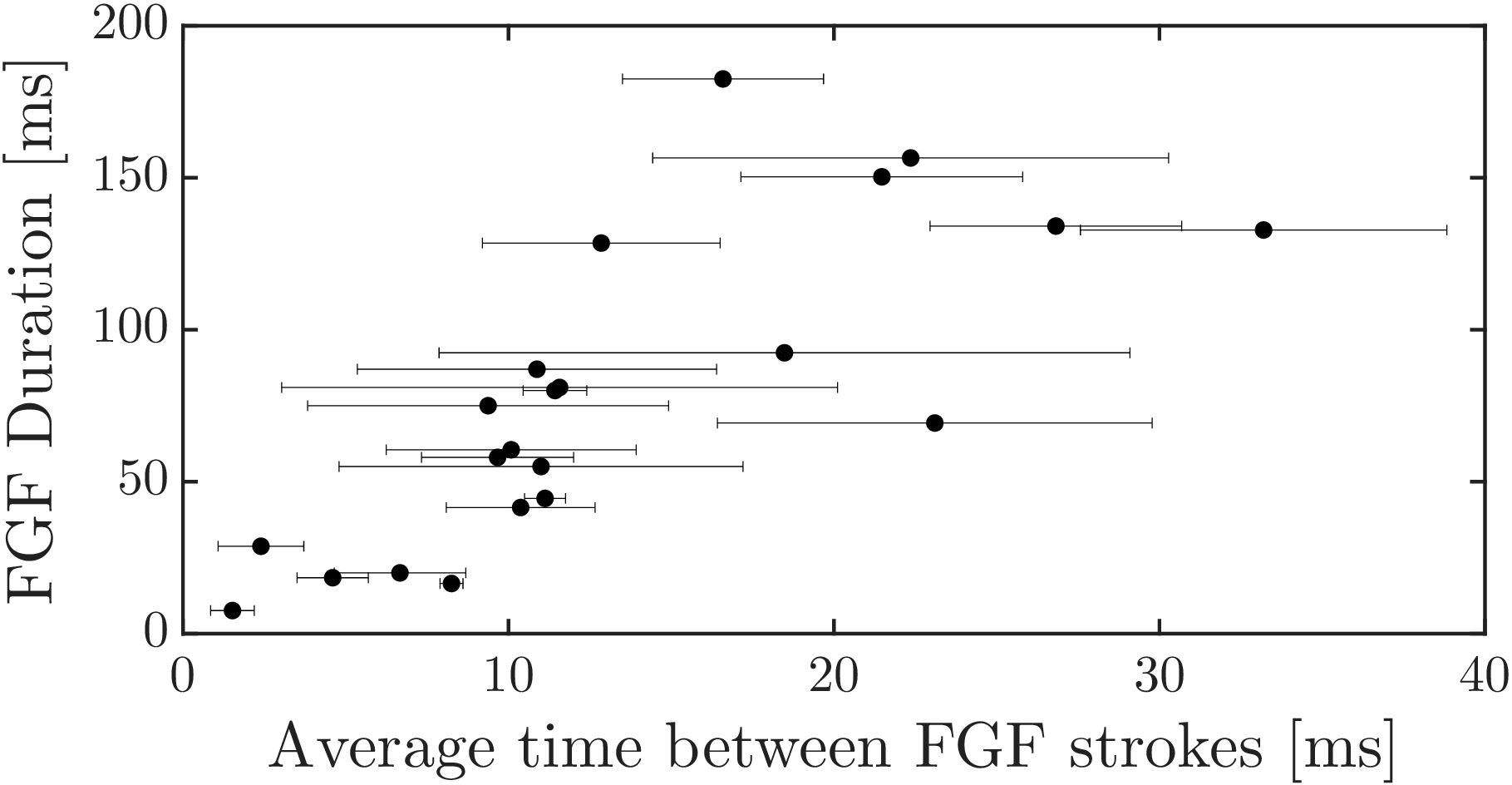}
  \caption{Plot summarizing the temporal characteristics of FGFs observed by iSTORM. The error bars represent the standard deviation of the sampled FGF pulses in the given FGF pulse train.}
  \label{fig:fgfTimeDiff}
\end{figure}

Fig.~\ref{fig:fgfTimeDiff} plots the FGF duration, marked as the time difference between the first detected pulse and the last, vs. the average time differences between pulses for a given FGF. The plot shows a relationship between the two characteristics, giving insight into the underlying electric conditions within the cloud. The error bars in \textit{x} represent the standard deviation in time between pulses for a given FGF. This exhibits that FGF pulses do not maintain a constant tempo but have a distribution in them. Perhaps more significant is the non-detection of optical or radio counterparts to the FGFs, indicating that it is possible that the only remote sensing signal originating from the underlying phenomena may be these high-energy phenomena.

\section{Conclusion}

We developed a gamma-ray instrument called the in-Situ Thunderstorm Observer for Radiation Mechanisms (iSTORM) that was deployed as part of the 2023 ALOFT campaign on a NASA ER-2 aircraft. The instrument contains 32 $\mathrm{CeBr}_3$ detectors along with a plastic scintillator and a bare SiPM. The campaign has proven very productive in observing a variety of atmospheric gamma-ray transients, including glows and TGFs. In addition, a new phenomenon known as flickering gamma-ray flashes has been observed for the first time.

The iSTORM instrument discussed in the manuscript is being upgraded by replacing the BeagleBone Black with a Raspberry Pi, which can collect the data from the front end at a faster rate. Next, we developed a new capability to send out health, status, and count rate data to the host aircraft, which will then be telemetered to the ground for near-real time gamma-ray flux analysis. The upgraded iSTORM instrument is slated to fly on the ER-2 as part of the 2026/2027 INSPYRE campaign, a mission to study pyrocumulonimbus systems originating from wildfires~\cite{INSPYRE, remington2025investigations}.

A new instrument, called iSTORM-CAT (Coded Aperture Telescope), is being developed for future campaigns~\cite{marisaldi2026imaging}. It's an imaging spectrometer that leverages a coded-aperture mask for imaging of the gamma rays. It will contain an array of CsI scintillators as well as a tantalum-based random-coded mask.

\section*{Acknowledgment}

This work is supported by the Office of Naval Research 6.1.

The ALOFT campaign was supported by the European Research Council under the European Union's Seventh Framework Programme (FP7/2007–2013)/ERC grant agreement no. 320839 and the Research Council of Norway under contracts 223252/F50(CoE) and contract 325582. The authors thank the NASA ER-2 Project Team at NASA Armstrong Flight Research Center for supporting the ALOFT campaign, and the MacDill Air Force Base for hosting. Significant financial and logistical support for ALOFT was provided by the NASA Earth Science Division.

%

\bibliographystyle{elsarticle-num}

\bibliography{IEEEbib} 
\end{document}